\documentclass[12pt]{article}
\usepackage[dvips]{graphicx}
\usepackage{cite}

\def\vect#1{{\mbox{\boldmath $#1$}}}
\def\+{\mbox{\unboldmath $+$}}
\def\Eo{\mathcal{E}^{\mathrm{odd}}}
\def\Ee{\mathcal{E}^{\mathrm{even}}}
\def\vr{\mbox{\boldmath$r$}}
\def\tz{{t_3}}
\def\ttz{{tt_3}}

\def\hh{{ij}}

\def\ss{{\sigma\sigma'}}
\def\tt{{\tau\tau'}}
\def\HF{\mathrm{HF}}
\def\RPA{\mathrm{RPA}}
\def\vrho{\vect{s}}

\begin{document}

\title{Mixed Representation RPA Calculation for Octupole Excitations 
on Superdeformed Sates in the $^{40}$Ca and Neutron-Rich Sulfur Regions}
\author{T. Inakura,$^a$ H. Imagawa,$^b$ Y. Hashimoto,$^b$\\
S. Mizutori,$^c$ M. Yamagami,$^d$ and K. Matsuyanagi$^e$\\ 
{\small\it $^a$Graduate School of Science and Technology, 
           Niigata University, Niigata 950-2181, Japan}\\
{\small\it $^b$Institute of Physics, University of Tsukuba, 
           Ibaraki 305-8571, Japan}\\ 
{\small\it $^c$Department of Human Science, Kansai Women's College, 
           Osaka 582-0026, Japan}\\ 
{\small\it $^d$Radioactive Isotope Physics Laboratory, RIKEN, 
           Saitama 351-0198, Japan}\\
{\small\it $^e$Department of Physics, Graduate School of Science, 
           Kyoto University, Kyoto 606-8502, Japan} }
\maketitle
\begin{abstract}
By means of the mixed representation RPA based on the Skyrme-Hartree-Fock
mean field, we investigate low-frequency octupole excitations 
built on the superdeformed (SD) states in the $N=Z$ nuclei around $^{40}$Ca 
and the neutron-rich Sulfur isotopes. 
The RPA calculation is carried out fully self-consistently
on the three-dimensional Cartesian mesh in a box, 
and yields a number of low-frequency octupole vibrations 
built on the SD states in $^{32}$S, $^{36}$Ar, $^{40}$Ca and $^{44}$Ti. 
In particular, a strongly collective $K^\pi=1^-$ octupole vibration is 
suggested to appear on top of the SD state in $^{40}$Ca.
For $^{48,50}$S close to the neutron drip line,
we find that the low-lying state created by the excitation 
of a single neutron from a loosely bound low $\Omega$ level
to a high $\Omega$ resonance level acquires an extremely strong 
octupole transition strength due to the spatially very extended 
structure of the particle-hole wave functions.
\end{abstract}

\section{Introduction}

Nowadays, more than two hundreds superdeformed (SD) bands are identified 
in various mass ($A$=60, 80, 130, 150, 190) regions
\cite{nol88,abe90,jan91,bak95,bak97,dob98}.
Every SD regions have their own characteristics so that we can significantly 
enlarge and deepen our understanding of nuclear structure by systematically
investigating similarities and differences among the SD bands 
in different mass region.
The SD shell structure is significantly different from 
that of normal deformation.
Namely, each major shell at the SD shape consists of about equal numbers of 
positive- and negative-parity levels.
This is a favourable situation for the appearance of
negative-parity collective modes. 
In fact, various mean-field calculations \cite{dud90,ska92,but96} 
and quasiparticle random phase approximation (RPA) \cite{miz91,miz93} 
on the basis of the rotating mean field (cranked shell model) indicated that 
SD nuclei are very soft against both 
the axial and non-axial octupole deformations. 
Thus, low-frequency octupole vibrations have been predicted 
to appear near the SD yrast lines \cite{nak96}, 
and recently discovered in experiments for heavy SD nuclei 
in the Hg-Pb region \cite{kor01}, 
and also in $^{152}$Dy \cite{lau02}.

In recent years, the SD bands have been discovered 
also in the $^{40}$Ca region:  
In $^{36}$Ar the SD band has been identified up to its termination 
at $I^\pi=16^+$ \cite{sve00,sve01a,sve01b}.
The SD band in the spherical doubly magic nucleus $^{40}$Ca is built 
on the 8p-8h excited $0^+$ state at 5.213 MeV \cite{ide01,chi03}.
The SD shell gap at $N=20$ ($Z=20$) is associated with 
the neutron (proton) 4p-4h excitation 
from below the $N=20$ ($Z=20$) spherical closed shell to the $f_{7/2}$ shell.
The rotational band built on the excited $0^+$ state
at 1.905 MeV in $^{44}$Ti may also be regarded as belonging to a family
of the SD band \cite{lea00}.
The fact that rotational bands built on excited
$0^+$ states are systematically observed is a unique feature of 
the SD bands in the $^{40}$Ca region, 
considering that the low angular momentum portions of the SD bands 
in heavy nuclei are unknown in almost all cases
(except the fission isomers). 
We have confirmed that the symmetry-unrestricted Skyrme-Hartree-Fock (SHF)
calculation indeed yields the SD local minima corresponding to these
experimental data \cite{ina02}.
It should be stressed that, in spite of the remarkable progress in experiment,
the doubly magic SD band in $^{32}$S associated with the SD magic number
$N$=$Z$=16, which has been anticipated quite a long time 
\cite{she72,lea75,rag78,ben81,gir83,yam00,mol00,rod00,tan01,afa00},
has not yet been observed and remains as a great challenge.

If we believe that $Z$=16 is a good SD magic number,
the existence of the SD shell gap at $N=20$, 
revealed by the discovery of the SD band in $^{40}$Ca \cite{ide01},
suggests that another SD band would appear 
in the neutron-rich nucleus $^{36}$S.
Furthermore, combining with the fact that the SD bands have been observed
in $^{60,62}$Zn \cite{sve99,sve97}, we can expect that 
Sulfur isotopes around $^{48}$S,
which are situated close to the neutron-drip line \cite{wer94,wer96},
constitute a new SD region associated with the SD shell gaps 
at $Z=16$ and $N \simeq 32$.
In fact, the symmetry-unrestricted SHF calculation \cite{ina03a} 
yields the SD local minima for $^{48,50}$S as well as $^{32}$S.

The investigation of low-frequency octupole vibrations built on
the SD states in the $A$=30-50 region possesses some new features
that are absent in the study of heavy SD nuclei.
For the $N=Z$ nuclei in the $^{40}$Ca region, 
it may be possible to observe in experiment such collective modes 
built on the known SD $0^+$ states. 
Moreover, because the proton and neutron shell structures are essentially 
the same, we can expect that strong coherence takes place between the
proton and neutron excitations and brings about an enhanced 
collectivity of these modes.
Concerning the anticipated new SD region around $^{48}$S,
we will encounter an essentially new situation: Because these
nuclei are situated close to the neutron drip line, 
there is almost no bound state above the Fermi surface and 
nucleons may excite into the continuum states even 
in the lowest excited states.

The study of soft collective modes unique to
unstable nuclei close to the neutron drip line is one of 
the current major subjects in nuclear structure physics.
For this purpose, the continuum RPA \cite{ber75,shl75} 
based on the Green's function method 
correctly treating the continuum boundary condition has been widely
used (see, e.g., \cite{ham99,sag01}).
Recently, this method has been extended \cite{mat01,kha02,yam04} 
to include the pairing correlations on the basis of 
the Hartree-Fock-Bogoluibov theory.
It is called the continuum quasiparticle-RPA (QRPA).
Furthermore, a new technique of solving the QRPA equations using the canonical 
basis has been developed \cite{ter05}. 
The canonical basis has also been used in the
QRPA calculation based on the relativistic mean field scheme 
\cite{vre01,paa05}.
A serious limitation of these works is that, 
because the construction of the Green's function for deformed mean fields 
is difficult, they are restricted to spherical nuclei.
Quite recently, however, this limitation has been overcome by 
Nakatsukasa and Yabana \cite{nak05}; they have proposed an iterative method 
of constructing the response functions for deformed systems with 
the proper boundary condition in the three-dimensional (3D) coordinate space.
They have also proposed a feasible method of 
treating the continuum boundary condition 
by means of the real-space time-dependent HF approach 
with the absorbing boundary condition.

In this paper, we investigate low-frequency octupole excitations
built on the SD states in neutron rich Sulfur isotopes as well as those
in the $^{40}$Ca region. 
For this purpose, we employ the mixed representation RPA 
\cite{lem68,ber83,mut02} 
based on the SHF mean field.
In this RPA scheme, the {\it particles} above the Fermi surface are described 
in the coordinate representation, 
while the {\it holes} are represented in the HF single-particle basis.
The RPA calculation is carried out fully self-consistently
on the 3D Cartesian mesh in a box. 
The major advantage of this approach is that it is a fully self-consistent
scheme in the sense that the same effective interaction is used in both 
the mean field and the RPA calculations;
i.e., all terms of the Skyrme interaction contributing to 
the RPA equations of motion are taken into account. 
It should be stressed here that such a fully self-consistent calculation
using the Skyrme-type interaction is difficult in the Green's function 
approach, so that, usually, the residual particle-hole interactions 
associated with the spin-orbit and Coulomb interactions, 
as well as those associated with the time-odd components
(related to the spin and current densities) of the mean field,
are ignored in the continuum RPA based on the SHF mean field. 
Another important merit of our approach is that, 
thanks to the use of the 3D Cartesian mesh representation,
we can treat strongly deformed nuclei on the same footing as spherical nuclei.
Furthermore, because the {\it particles} are described in the coordinate 
representation, we do not need to introduce an upper cut off for their
energies.
Quite recently, Imagawa and Hashimoto 
constructed a new computer code that carries out the 
fully self-consistent RPA calculation in the mixed representation 
on the basis of the SHF mean field, and they carefully tested its 
numerical accuracy \cite{ima03a,ima03b,ima05}. 
The numerical calculations of this paper were carried out
using a refined version this code.

This paper is organized as follows:
In Section 2, a brief account of the self-consistent RPA calculation 
using the mixed representation is given. 
In Section 3, we present and discuss the results of numerical calculation 
for low-frequency octupole vibrations built on the SD states 
in $N=Z$ nuclei, $^{32}$S, $^{36}$Ar, $^{40}$Ca, and $^{44}$Ti.
In Section 4, the results for neutron-rich Sulfur 
isotopes, $^{36}$S, $^{48}$S, and $^{50}$S, are presented and discussed. 
The main results of this paper are summarized in Section 5.

A preliminary version of this work was reported in 
Refs.~\cite{ina04,ina05}.

\section{RPA calculation using the mixed representation}

\subsection{Basic formulae}

The RPA equations in the mixed representation are easily derived
either by means of the lineralized equations of motion approach \cite{lem68}
or the small-amplitude approximation of the time-dependent HF theory
\cite{ber83}. Here, we recapitulate the basic formula 
following the former approach.

The creation and annihilation operators of nucleon,
$\psi^\dag(x)$ and $\psi(x)$, are divided into those for
the bound states and those for the continuum states:  
\begin{eqnarray}
\psi^\dag(x) = \sum_\alpha \varphi^\ast_\alpha(x) c^\dag_\alpha 
+ \int_0^{\infty}  \mathrm{d}\varepsilon  \varphi^\ast_\varepsilon (x) 
c^\dag_\varepsilon,\\
\psi (x) = \sum_\alpha \varphi_\alpha(x) c_\alpha 
+ \int_0^{\infty}  \mathrm{d}\varepsilon  
\varphi_\varepsilon (x) c_\varepsilon, 
\label{cre_ord}
\end{eqnarray}
where $x$ denotes a set of space, spin and isospin coordinates,
i.e., $x=\{\vr, \sigma, \tau\}$.
The symbol $\alpha$ denotes the HF bound states with
wave functions $\varphi_\alpha(x)$, while the energy $\varepsilon$  
specifies the continuum states with wave functions $\varphi_\varepsilon(x)$.
With the use of the {\it particle} and {\it hole} concepts, we then divide
the nucleon operators according to the occupation number 
$\theta_\alpha$ in the HF ground state 
($\theta_\alpha=0$ for unoccupied states, 
$\theta_\alpha=1$ for occupied states).
We consider even-even nuclei and assume that the single-particle states
$\alpha$ and their time reversed states are doubly degenerated.
Obviously, the continuum states belong to the particle space.
On the other hand, for the bound states,
the operators ($c^\dag_\alpha$, $c_\alpha$) 
are divided into the particle operators ($a^\dag_m$, $a_m$) 
and the hole operators ($b^\dag_i$, $b_i$),
\begin{eqnarray}
c^\dag_\alpha = (1-\theta_\alpha) a^\dag_m + \theta_\alpha b_i, \\
c_\alpha = (1-\theta_\alpha) a_m + \theta_\alpha b^\dag_i,
\end{eqnarray}
where the indices $m$ and $i$ are used in place of $\alpha$ to distinguish 
the particle and hole states. 
Thus, the particle creation and annihilation operators,
$a^\dag(x)$ and $a(x)$, at the coordinate $x$ are written 
in terms of the above ($a^\dag_m$, $a_m$) and
the integration with respect to the positive energy $\varepsilon$
specifying the continuum states; 
\begin{eqnarray}
a^\dag(x) = \sum_m \varphi^\ast_m(x)  a^\dag_m + \int  \mathrm{d}
\varepsilon  \varphi^\ast_\varepsilon(x)  a^\dag_\varepsilon ,\\ 
a(x) = \sum_m \varphi_m(x)  a_m + \int  \mathrm{d}
\varepsilon  \varphi_\varepsilon(x)  a_\varepsilon.
\end{eqnarray}
In the mixed representation, 
the coordinate representation is used for particles, 
while the HF basis specified by the discrete index $i$ are used for holes.
Thus, the following expressions of the nucleon operators are convenient:
\begin{eqnarray}
\psi^\dag(x) = a^\dag(x) + \sum_i \varphi^\ast_i(x) b_i, \\
\psi(x) = a(x) + \sum_i \varphi_i(x) b_i^\dag.
\label{cre_mix}
\end{eqnarray}
Note that 
\begin{equation}
\{ a(x), a^\dag(x') \} = P(x, x') = \delta(x,x') - Q(x,x') 
\equiv  \delta(x,x') -  \sum_i \varphi_i(x) \varphi^\ast_i(x'),
\end{equation}
where $P(x, x')$ and $Q(x, x')$ are the projectors onto the
particle and hole spaces, respectively.

Using an abbreviation, 
$\sum_x \equiv \sum_{\sigma\tau} \int  \mathrm{d}\vr$, 
the RPA phonon creation operators in the mixed representation are written as
\begin{eqnarray}
O^\dag_\lambda = \sum_i \sum_x \left\{ X^\lambda_i(x)  a^\dag(x) b^\dag_i
- Y^\lambda_i(x)  b_i  a(x) \right\} .
\end{eqnarray}
Then, the RPA eigenvalue equations determining the eigen-energies, 
$\hbar \omega_\lambda$, and the forward and backward amplitudes,
$X_i^\lambda(x)$ and $Y_i^\lambda(x)$ are given in the following form:
\begin{eqnarray}
\sum_{j}\sum_{x'} \left [ A_{\hh}(x,x') X_j^\lambda(x') + 
B_{\hh}(x,x') Y_j^\lambda(x') \right ] 
&=& \hbar \omega_\lambda X_i^\lambda(x) 
\label{RPAequone}\\ 
\sum_{j}\sum_{x'} \left [ B_{\hh}^*(x,x') X_j^\lambda(x') + 
A_{\hh}^*(x,x') Y_j^\lambda(x') \right ] 
&=& - \hbar \omega_\lambda Y_i^\lambda(x),
\label{RPAequtwo}
\end{eqnarray}
where
\begin{eqnarray}
A_{\hh}(x,x') = \sum_{x'',x'''} P(x,x'') \tilde{A}_{\hh}(x'',x''') P(x''',x'),\\
B_{\hh}(x,x') = \sum_{x'',x'''} P(x,x'') \tilde{B}_{\hh}(x'',x''') Q(x''',x'),
\end{eqnarray}
and
\begin{eqnarray}
\tilde{A}_{\hh}(x,x') &=& \left[ h_\HF(x,x') - e_i  \delta(x,x') \right]
\delta_{\hh}
+ \sum_{x''x'''} \varphi_i(x'')\varphi^\ast_j(x''') \frac{\partial^2 E[\rho]}
{\partial \rho(x'',x) \partial \rho(x',x''')} ,\\
\tilde{B}_{\hh}(x,x') &=& \sum_{x''x'''} \varphi_i(x'')\varphi_j(x''') 
\frac{\partial^2 E[\rho]}{\partial \rho(x'',x) \partial \rho(x''',x')}. 
\end{eqnarray}
The above RPA equations can be recast into the following form
for the linear combinations, 
$\phi^{(\pm)\lambda}_i(x) = X^\lambda_i(x) \pm Y^{\lambda\ast}_i(x)$:
\begin{eqnarray}
\sum_{x'} \left[ h_\HF(x,x') - e_i  \delta(x,x') \right] 
\phi^{(+)\lambda}_i(x') + \sum_{x'x''} P(x,x') h^{(+)\lambda}_\RPA(x',x'') 
\varphi_i(x'') = \hbar \omega_\lambda \phi^{(-)\lambda}_i(x) \ , 
\label{RPAfinalp}\\
\sum_{x'} \left[ h_\HF(x,x') - e_i  \delta(x,x') \right] 
\phi^{(-)\lambda}_i(x') + \sum_{x'x''} P(x,x') h^{(-)\lambda}_\RPA(x',x'') 
\varphi_i(x'') =  \hbar \omega_\lambda \phi^{(+)\lambda}_i(x) \ .
\label{RPAfinalm}
\end{eqnarray}
Here, the RPA Hamiltonian is given by 
\begin{eqnarray}
h^{(\pm)\lambda}_\RPA(x,x')  = 
\sum_{x'',x'''} \rho^{(\pm)\lambda}(x'',x''') 
\frac{\delta h_\HF(x,x')}{\delta \rho(x'',x''')}, 
\label{RPA_Hamiltonian}
\end{eqnarray}
where
\begin{eqnarray}
\rho^{(\pm)\lambda}(x,x') = \sum_i 
\left[ \varphi^\ast_i(x') \phi^{(\pm)\lambda}_i(x) 
\pm \phi^{(\pm)\lambda\ast}_i(x') \varphi_i(x)  \right]  .  
\label{def.transition}
\end{eqnarray}
represents the transition density.
The above equations are obtained also by making 
the small-amplitude approximation for the equations of motion of 
the time-dependent HF mean field \cite{ber83}.
The explicit expression of $h^{(\pm)\lambda}_\RPA(x,x')$ is given 
in Appendix. 
The transition matrix elements for the octupole operators $Q_{3K}$ 
are given by
\begin{eqnarray*}
\langle 0 | Q_{3K} | \lambda \rangle = \sum_x Q_{3K}(\vr) \sum_i \left[ X^\lambda_i(x) \varphi^\ast_i(x) + \varphi_i(x) Y^\lambda_i(x) \right].
\end{eqnarray*}

\subsection{Details of numerical calculation}

We solve the SHF equation on the 3D Cartesian mesh 
assuming the reflection symmetry about
the $(x,y)$-, $(y,z)$-, and $(z,x)$-planes. 
The derivative operators are treated by means of 
the Lagrange mesh method\cite{bay86}, 
except for the derivative of the Coulomb potential 
for which the finite difference method with the 9 points formula is used.
Based on the numerical test explained below,
we adopt the mesh spacing $\Delta x = 0.6$ fm. 
Because we treat the superdeformed nuclei, 
we use the rectangular box with 15 and 25 mesh points 
in the positive directions of 
the minor and major axes, respectively, i.e., 
we take the mesh points at 
$x=0.3, 0.9, ...,8.7$ fm, $y=0.3, 0.9, ...,8.7$ fm and 
$z=0.3, 0.9, ...,14.7$ fm (the major axis is called the $z$-axis). 
Considering the reflection symmetries, the total size of the box is thus
$17.4 \times 17.4 \times 29.4$ fm$^3$.
For the effective interaction, the standard versions of the Skyrme interaction, 
SIII~\cite{bei75}, SkM$^\ast$~\cite{bar82} and SLy4~\cite{cha97},
are used. 

We then solve the RPA equations, (\ref{RPAfinalp}) and (\ref{RPAfinalm}), 
using the SHF solutions obtained above. 
With the mesh discretization, 
Eqs.~(\ref{RPAfinalp}) and (\ref{RPAfinalm}) become 
the matrix eigenvalue problem with the $2\times N_h\times N_p$ 
dimension, where $N_h$ and $N_p$ denote the numbers of the hole states and  
the mesh points, respectively.  Because we assume the reflection symmetry 
about the three planes, the RPA eigen-value equations are 
separated into four blocks specified by the parity quantum numbers. 
It is easily ascertained that the negative parity operators are classified 
into the four groups,
$(-,+,+)$, $(+,-,+)$, $(+,+,-)$, and $(-,-,-)$, 
where the first, the second, and the third signs indicate the parity 
with respect to the reflection about 
the $(y,z)$-, $(z,x)$-, and $(x,y)$-planes, respectively.
For instance, the octupole operator 
$Q_{30} = r^3Y_{30}$ belongs to the $(+,+,-)$ sector. 
On the other hand, the octupole operators with $K\ne0$,  
$Q_{3K}^{(\pm)} = r^3 \left ( Y_{3K} \pm Y_{3,-K} \right )/\sqrt{2}$,
are classified according to the parity quantum numbers  
as shown in Table \ref{table:symmetry}.
We calculate the eigenvalues and eigenvectors
of the RPA matrix by means of the conjugate gradient method
in a version developed in Refs. \cite{ima03a,ima03b, ima05}. 

\begin{figure}[p]
\begin{center}
\includegraphics[width=0.50\textwidth,keepaspectratio]
{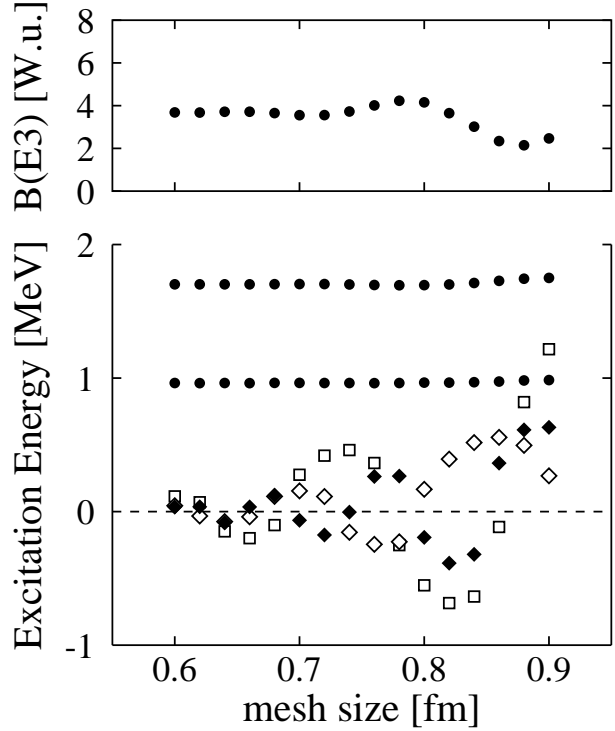}
\caption{{\small
Low-energy solutions of the RPA for negative-parity excitations on the
SD ground state in $^{40}$Ca, plotted as functions of the mesh spacing
used in the SHF+RPA calculation with the SIII interaction. 
{\it Lower part}:  
Energies of the spurious center of mass modes belonging to
the $(-,+,+)$, $(+,-,+)$ and $(+,+,-)$  sectors are plotted 
with open diamonds, filled diamonds, and open squares, respectively. 
When their excitation energies, $\hbar\omega_\lambda$, take imaginary values, 
the $-|\omega_\lambda|$ values are plotted for convenience of presentation. 
The filled circles at about 1.0 and 1.7 MeV indicate the lowest physical 
excitations in the $(-,+,+)$ and $(+,-,+)$ sectors, respectively. 
{\it Upper part}:  
Transition strengths for the lowest physical excitation at about 1.7 MeV 
in the $(+,-,+)$ sector, plotted as a function of mesh spacing.
}}
\label{check.span.SD}
\end{center}
\end{figure}

It should be noted here that the translational symmetry is broken 
in the coordinate mesh representation so that the eigenvalues of the
RPA corresponding to the spurious center of mass motion do not necessarily
become zero. It is possible that the eigenvalue $\hbar\omega_\lambda$ 
corresponding to the spurious mode become pure imaginary.
Moreover, in deformed nuclei, except for the $(-,-,-)$ sector, 
these spurious modes can mix with the octupole modes 
having the same symmetry properties
(see Table \ref{table:symmetry} for the classification of 
the dipole operators corresponding to the center of mass motion).
It is therefore very important to carefully check the accuracy of numerical
calculation whether or not the spurious center of motions 
are decoupled from the physical excitations in a good approximation.
In the numerical calculation, we manipulate this problem with the
following procedure.
Combining Eqs. (\ref{RPAequone}) and (\ref{RPAequtwo}),
the RPA eigenvalue problem is transformed into the following form:
\begin{eqnarray}
\sum_{jk}\sum_{x',x''}
[A_{ij}(x,x')A_{jk}(x',x'') &-&
 A_{ij}(x,x')B_{jk}(x',x'') \nonumber \\ 
+B_{ij}(x,x')A_{jk}(x',x'') &-&
 B_{ij}(x,x')B_{jk}(x',x'') 
 ] \phi_k^{(-)\lambda}(x'') = (\hbar\omega_\lambda)^2
 \phi_i^{(-)\lambda}(x). \label{RPAspurious}
\end{eqnarray}
This form is convenient for applying the conjugate gradient method
which is valid only for real eigenvalues,
because, when the eigenvalue $\hbar\omega_\lambda$ of the spurious mode
take an imaginary value, we can easily identify it  
as a negative $(\hbar\omega_\lambda)^2$ solution of this equation. 
The result of such an accuracy test of numerical calculation 
is presented in Fig.~\ref{check.span.SD}.
In this figure, low-energy solutions of the above RPA equation 
for negative-parity excitations on the SD ground state in $^{40}$Ca
are plotted as functions of the mesh spacing  $\Delta x$
used in the SHF-RPA calculation with the SIII interaction. 
We see that a good convergence is attained at about $\Delta x =0.6$ fm,
although excitation energies of the spurious center of mass modes 
fluctuate in the region of $\Delta x > 0.7$ fm. 
It should be noted that the excitation energies of the physical excitation 
modes and their transition strengths almost converge for $\Delta x < 0.7$ fm.
Thus, we adopt $\Delta x =0.6$ fm in the numerical calculations presented 
below. We note that the use of this mesh spacing corresponds the 
introduction of an effective energy cut-off 
$E_{\mathrm{cut-off}}=(\hbar^2/2m)({\pi}/\Delta x)^2 \sim 500$ MeV.  
This value is sufficiently large. Accordingly, it may be allowed to say 
that our calculation is, in practice, cut-off free.
Certainly, it is desirable to use a smaller $\Delta x$ and  a larger box 
for better numerical accuracy, but it is difficult to do so with the 
available computing power at the present time. 
This limitation will be overcome in the near future. 

In presenting the results of the SHF-RPA calculation in the subsequent 
sections, we label the SHF single-particle levels in terms of 
the asymptotic quantum numbers $[N_{\rm osc}, n_z, \Lambda,\Omega]$
for the modified oscillator potential.
It should be noted that this labeling is used solely for convenience
of presentation, and do not always mean that they are 
good quantum numbers. 
The labeling of the SHF single-particle states is done 
by examining their properties,
like reflection symmetries and expectation values of the angular momentum,
and by comparison with the single-particle level scheme 
obtained in Ref.~\cite{yos05}, 
where the deformed Woods-Saxon potential simulating the SHF potential 
was constructed and properties of both the bound states and 
the discretized continuum states for this potential are analysed
in detail. In this reference, the RPA calculations were also carried out
for the octupole excitations on the SD states under consideration,
using the conventional matrix formulation with these single-particle 
levels in a truncated model space and a density dependent contact
interaction. The results of this simple RPA calculation qualitatively
agree with the results of the mixed representation RPA calculation 
presented below. Thus, we have used the correspondence between the
two calculations for checking the numerical calculation.
As we shall discuss later, this comparison was especially useful 
for the purpose of distinguishing the resonant levels from 
the non-resonant discretized continuum levels.

\section{Results of calculation for the $^{40}$Ca region}

\begin{figure}[p]
\begin{center}
\includegraphics[width=0.80\textwidth,keepaspectratio]
{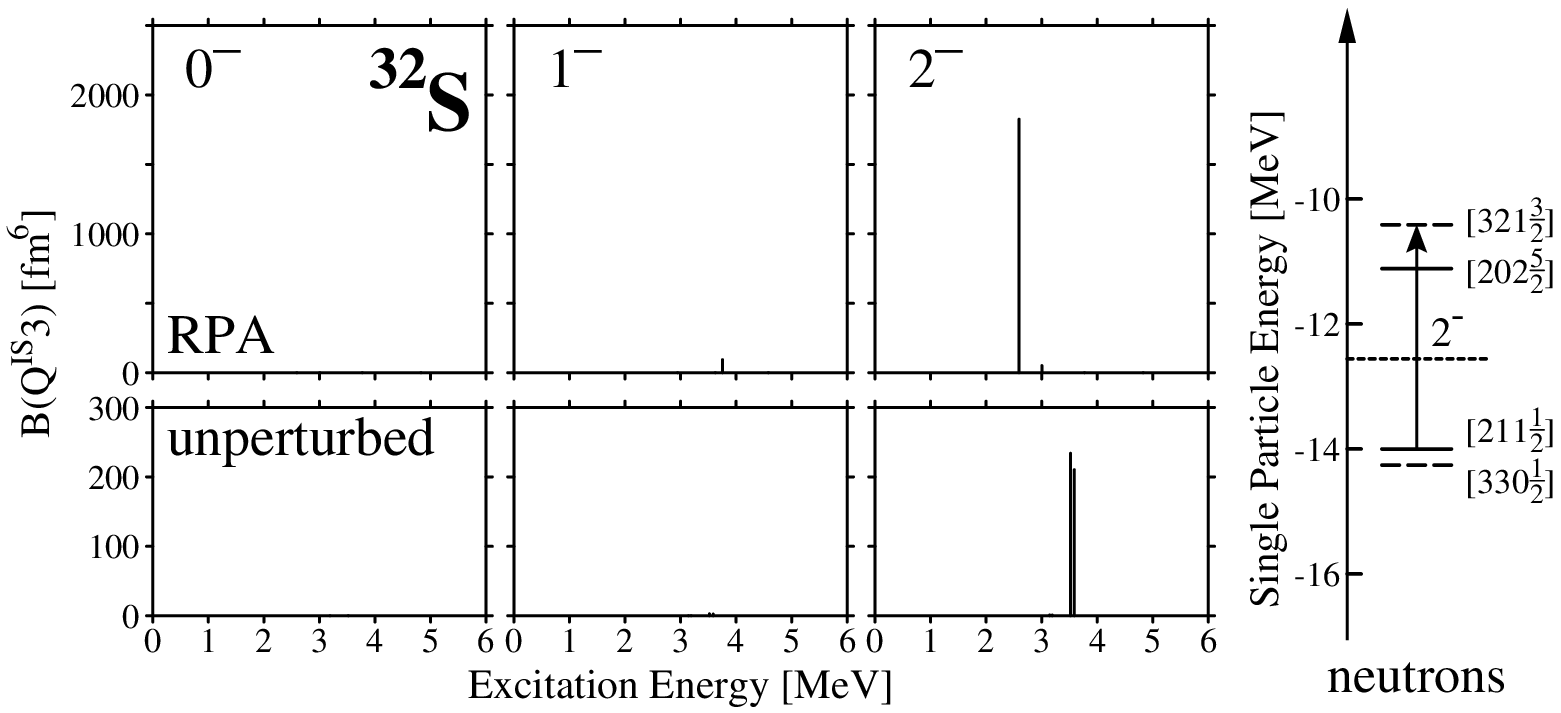}\\
\includegraphics[width=0.80\textwidth,keepaspectratio]
{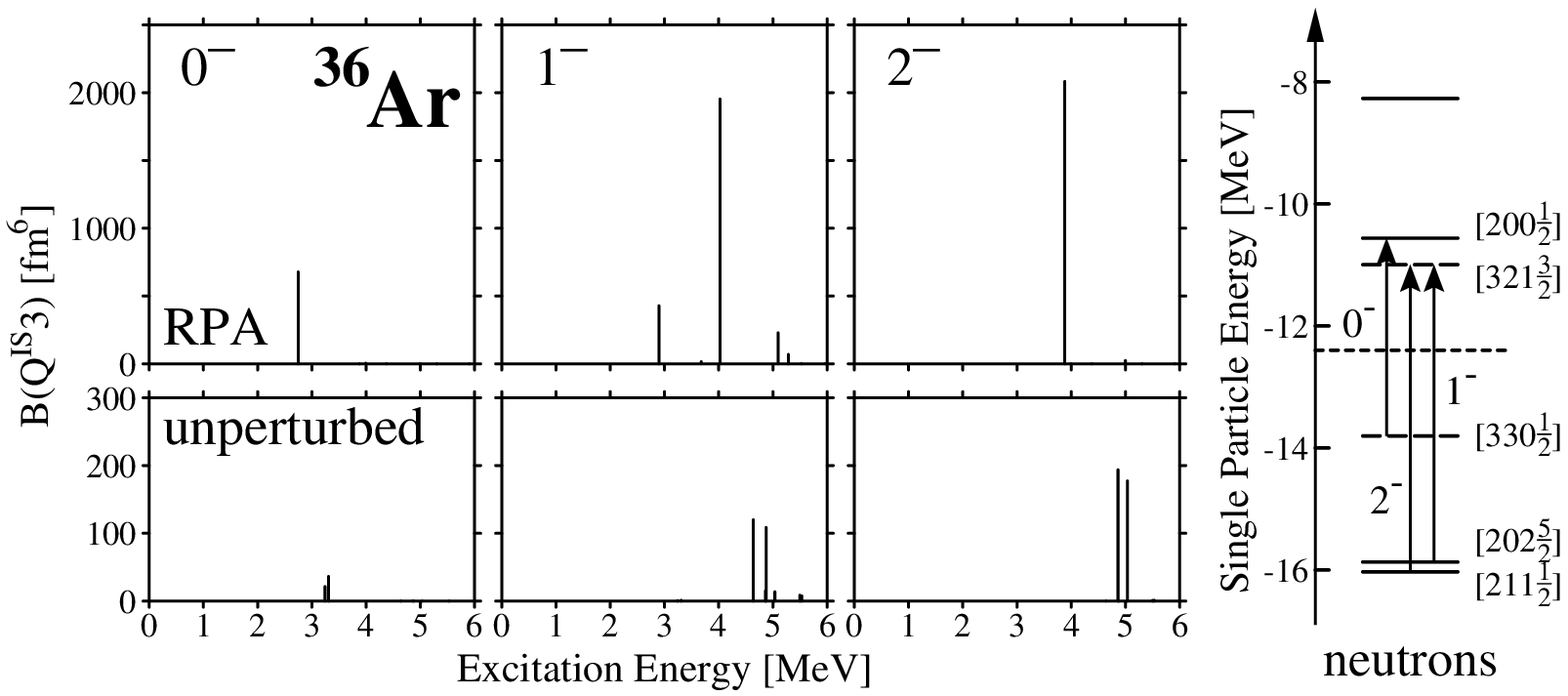}\\
\includegraphics[width=0.80\textwidth,keepaspectratio]
{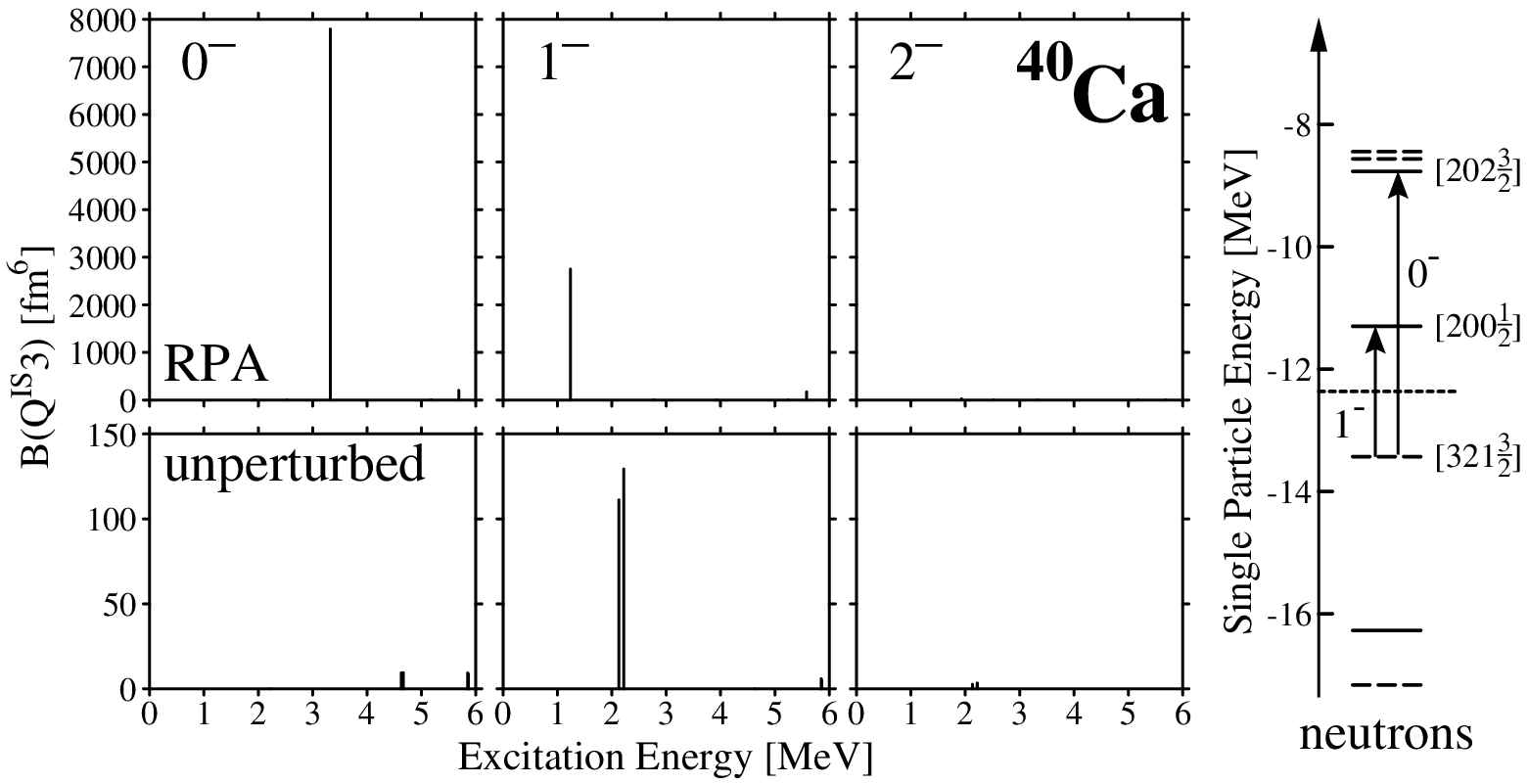}\\
\caption{{\small
Isoscalar octupole transition strengths for $K^{\pi}=0^{-},1^{-}$ and $2^{-}$ 
excitations on the SD states in (a) $^{32}$S, (b) $^{36}$Ar, and (c) $^{40}$Ca, 
obtained by the SHF-RPA calculation with the SIII interaction.
The unit is fm$^6$.
For comparison with the RPA strengths, unperturbed particle-hole strengths 
are also shown in the lower panels for each nucleus.
There are no significant strengths for $K^{\pi}=3^{-}$ excitations values 
in this low-energy region, so that they are not shown.
Some particle-hole excitations of neutrons near the Fermi surface 
are drawn by arrows with their $K^{\pi}$ values 
in the rightmost part for each nucleus.
We have similar excitations also for protons. 
The solid, dashed, and dotted lines indicate positive-parity levels, 
negative-parity levels, and the Fermi surface, respectively.
The asymptotic quantum numbers 
$[N n_z \Lambda \Omega]$ are indicated for pertinent levels.
}}
\label{RPA_unpert.NZ.SIII.SD.prolate}
\end{center}
\end{figure}

Figure~\ref{RPA_unpert.NZ.SIII.SD.prolate}  
displays the transition strength distributions obtained by 
the SHF-RPA calculation with the SIII interaction 
for low-frequency octupole excitations built on the SD states 
in $^{32}$S, $^{36}$Ar, and $^{40}$Ca.
The RPA transition stengths $B(Q^\mathrm{IS}3)$ are here defined 
as the squared matrix elements of the isoscalar octupole operators 
between the SD excited state $| \lambda \rangle$ 
and the SD ground state $| 0 \rangle$ , 
$B(Q^\mathrm{IS}3)=
 | \langle 0 | Q^\pi_{3K} \+ Q^\nu_{3K} | \lambda \rangle |^2 $. 
In this figure, the RPA transition strengths 
for the electric octupole operators, 
$B(E3)= | \langle 0 | Q^\pi_{3K} | \lambda \rangle |^2 $, 
are also indicated for levels possessing strong collectivity. 
Here, $Q^\pi_{3K}$ and $Q^\nu_{3K}$ denote the proton and neutron 
octupole operators. 
It should be noted that these quantities represent 
the intrinsic matrix elements squared.
Accordingly, in order to evaluate the reduced transition probabilities 
in the laboratory frame, 
appropriate Clebsch-Gordan (CG) coefficients squared and 
other kinematical factors should be multiplied to them \cite{bor75}.
Specifically, for the transition from the $3^-$ member of 
the $K=0$ ($K \ne 0$) band to the SD $0^+$ band head, 
the CG coefficient is unity and the kinematical factor is 1/7 (2/7).
The SHF calculation for $^{40}$Ca in fact yields a SD mean field with small
triaxially of $\gamma$=7$^\circ$, while the SD solutions for $^{32}$S
and $^{36}$Ar are axially symmetric.
Nevertheless, for simplicity, we here present the result of the RPA calculation 
on the basis of the SHF mean field obtained 
under the constraint $\gamma$=0$^\circ$. 
The effect of the small triaxial deformation 
on these excitation modes will be discussed in the end of this section.

For $^{32}$S, we obtain a low-frequency collective $K^\pi=2^-$ mode  
at 2.6 MeV above the SD band head. 
It possesses a large isoscalar octupole transition strength of 30 W.u.
(1 W.u. $\simeq$ 61 fm$^6$ for $^{32}$S).
The unperturbed strengths in the 3.5 MeV region, 
seen in Fig.~\ref{RPA_unpert.NZ.SIII.SD.prolate}, are associated with 
the proton and neutron excitations from the $[211\frac{1}{2}]$ 
level to the $[321\frac{3}{2}]$ level.  
The RPA transition strength is greatly enhanced
in comparison with the unperturbed strengths 
and the RPA eigen-energy is significantly shifted down from the unperturbed 
particle-hole excitation energies, so that 
this $K^\pi=2^-$ mode can be regarded as a collective octupole vibration.

For $^{36}$Ar, we obtain two strongly collective modes; 
a $K^\pi=2^-$ mode at 3.9 MeV and a $K^\pi=1^-$ mode at 4.0 MeV. 
They possess large isoscalar octupole transition strengths of 27 and 25 
W.u., respectively (1 W.u. $\simeq$ 77 fm$^6$ for $^{36}$Ar).
This $2^-$ mode is similar to that in $^{32}$S discussed above,
except for the following difference:   
Because the quadrupole deformation of the SD state in $^{36}$Ar 
is smaller ($\beta_2=0.51$) than that in $^{32}$S~($\beta_2=0.72$),
the unperturbed particle-hole excitation energies (4.9-5.0 MeV)
of protons and neutrons from the up-sloping $[211\frac{1}{2}]$ level 
to the down-sloping $[321\frac{3}{2}]$ level 
are larger in $^{36}$Ar than in $^{32}$S. 
Accordingly, the RPA excitation energy of the $2^-$ mode in 
$^{36}$Ar is higher than that in $^{32}$S. 
Concerning the $1^-$ mode,
its excitation energy is considerably shifted down 
from the particle-hole excitation energies (4.6-4.9 MeV)
of protons and neutrons
from the $[202\frac{5}{2}]$ level to the $[321\frac{3}{2}]$ level.
Considering this together with the enhanced transition strength,
this $1^-$ mode can also be regarded as a mode possessing 
strongly collectivity.
In addition to the $2^-$ and $1^-$ modes mentioned above,
we also obtain a $0^-$ mode and another $1^-$ mode 
in the 2.8-2.9 MeV region, which are moderately collective.

For $^{40}$Ca, 
we obtain a low-frequency collective $K^\pi=1^-$ mode at about 1.2 MeV.
It possesses a large isoscalar octupole transition strengths of 29 W.u. 
(1 W.u. $\simeq$ 95 fm$^6$ for $^{40}$Ca).
The unperturbed particle-hole strengths in the 2.2 MeV region 
are associated with the proton and neutron excitations 
from the $[321\frac{3}{2}]$ level to the $[200\frac{1}{2}]$ level.  
The fact that the RPA transition strength is significantly enhanced
in comparison with the unperturbed strength 
and the RPA eigen-energy is shifted down from the unperturbed 
particle-hole excitation energies indicates that 
this $K^\pi=1^-$ mode possesses a strong collectivity.
We also obtained a strongly collective $K^\pi=0^-$ mode 
possessing quite large strength (82 W.u.) at about 3.3 MeV. 
For this  $K^\pi=0^-$ transitions, 
although small unperturbed strengths (about 0.1 W.u.) 
associated with the particle-hole excitations of protons and neutrons 
from the $[321\frac{3}{2}]$ level to the $[202\frac{3}{2}]$ level
are seen at about 4.6 MeV, 
their strengths are small mainly because they strongly violate 
the asymptotic selection rule for low-energy octupole transitions 
in the SD harmonic-oscillator potential \cite{miz93}.
Thus, this collective $K^\pi=0^-$ mode is generated mainly from many
particle-hole configurations lying in the energy region higher
than that shown in this figure. 
We mention that these collective modes in $^{40}$Ca were 
previously reported by one of authors (H.I.) in Ref.~\cite{ima03a}.

For all the RPA eigen-modes 
presented in Fig. \ref{RPA_unpert.NZ.SIII.SD.prolate}, 
we can say the followings:\\
1) The approximate relation, $B(Q^\mathrm{IS}3) \approx 4 B(E3)$, holds, 
indicating that they are collective modes generated by coherent superposition 
of proton and neutron excitations.\\
2) The RPA strengths are greatly enhanced from the unperturbed strengths of 
individual particle-hole excitations. This indicates that these RPA eigen-modes 
are generated by collective contributions of quite a number of particle-hole 
excitations. \\
The role of coherence between the neutron and proton excitations 
in building up these collective enhancements can also be seen by comparing
these RPA transition strengths for the $N=Z$ nuclei, $^{32}$S and $^{40}$Ca, 
with those for the $N \ne Z$ nucleus $^{36}$S presented in 
the succeeding section.

\begin{figure}[p]
\begin{center}
\includegraphics[width=0.80\textwidth,keepaspectratio]
{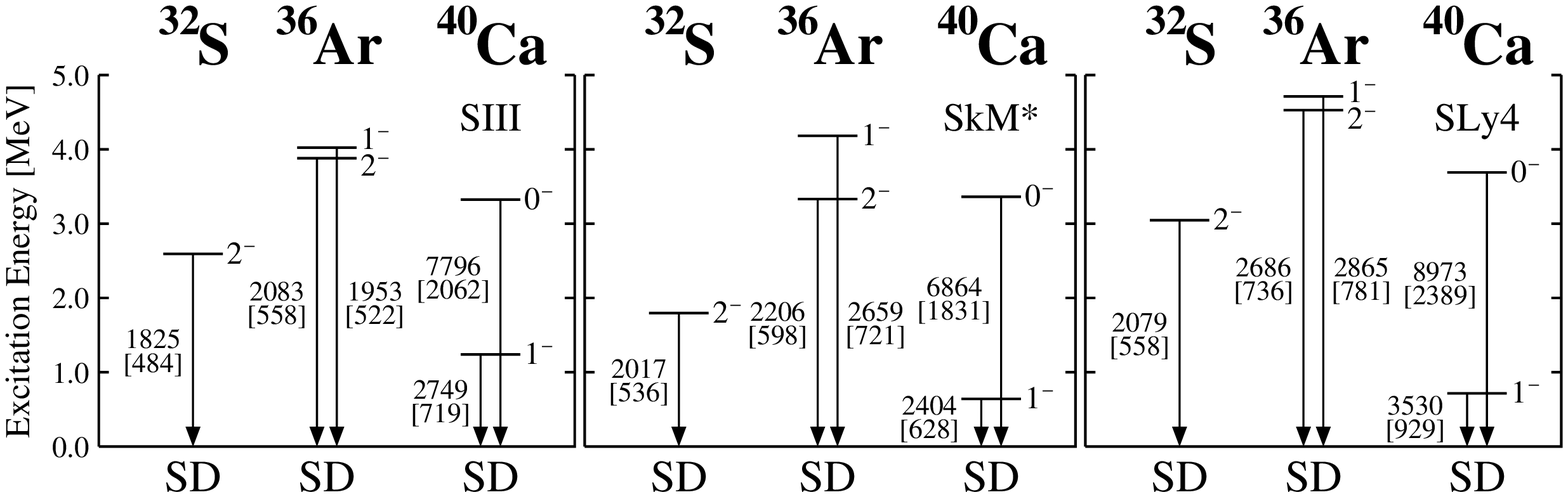}\\
\caption{{\small
Comparison of the SHF-RPA calculations with the SIII, SkM$^*$ and SLy4 
interactions for collective octupole excitations on the SD states 
in $^{32}$S, $^{36}$Ar, and $^{40}$Ca. 
Only collective excitations having $B(Q^{\rm IS}3)$ 
greater than 10 W.u. are displayed here 
(see Fig.~\ref{RPA_unpert.NZ.SIII.SD.prolate} 
for other excitations having smaller strengths).
The numbers on the right-hand sides of individual levels 
indicate the $K$ quantum numbers.
The numbers beside the arrows indicate the intrinsic transition matrix 
elements squared of the isoscalar octupole operator, $B(Q^\mathrm{IS}3)$, 
in units of fm$^6$.
Those for the electric octupole operator,
$B(E3)$, are also given in brackets in units of $e^2$fm$^6$.
}}
\label{rpa.NZ.prolate}
\end{center}
\end{figure}

We also carried out the same kind of calculation 
using the SkM$^\ast$\cite{bar82} and the SLy4\cite{cha97} interactions.
The results for strongly collective excitations 
of these SHF-RPA calculations with different versions of the
Skyrme interaction are compared in Fig.~\ref{rpa.NZ.prolate}.
Note that the calculations for 
$^{40}$Ca were done with the constraint $\gamma$=0$^\circ$ for the mean fields, 
although the SHF calculations yield small triaxial deformations; 
4$^\circ$ and 8$^\circ$ with the SkM$^\ast$ and SLy4 interactions, 
respectively. For all excitation modes discussed above,
we obtained the results qualitatively similar to those 
with the SIII interaction,  
Namely, our prediction about the existence of these collective modes 
is rather robust, depending little on the choice of the
Skyrme interaction.

\begin{figure}[p]
\begin{center}
\includegraphics[width=0.80\textwidth,keepaspectratio]{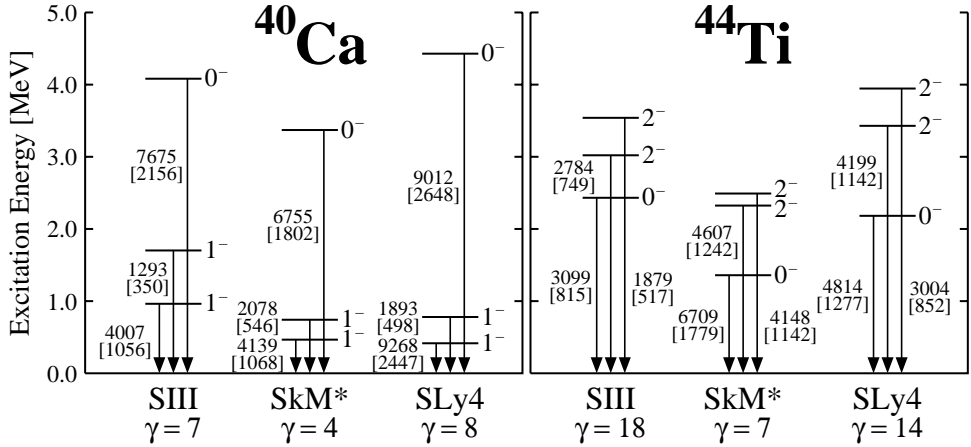}
\caption{{\small 
Low-frequency collective octupole excitations on the SD states in $^{40}$Ca 
and $^{44}$Ti, obtained by the SHF-RPA calculation taking into account
the triaxial deformation of the mean field.
Only collective excitations having $B(Q^{\rm IS}3)$ 
greater than 10 W.u. are displayed here. 
The numbers on the right-hand sides of individual levels 
indicate the approximate $K$ quantum numbers,
The numbers beside the arrows indicate the intrinsic transition matrix 
elements squared of the isoscalar octupole operator, $B(Q^\mathrm{IS}3)$, 
in units of fm$^6$.
Those for the electric octupole operator,
$B(E3)$, are also given in brackets in units of $e^2$fm$^6$.
The Skyrme interaction used and the triaxial deformation $\gamma$ 
of the mean field are indicated at the bottom.
}}
\label{rpa.NZ.triaxial}
\end{center}
\end{figure}

\begin{figure}[p]
\begin{center}
\includegraphics[width=0.80\textwidth,keepaspectratio]
{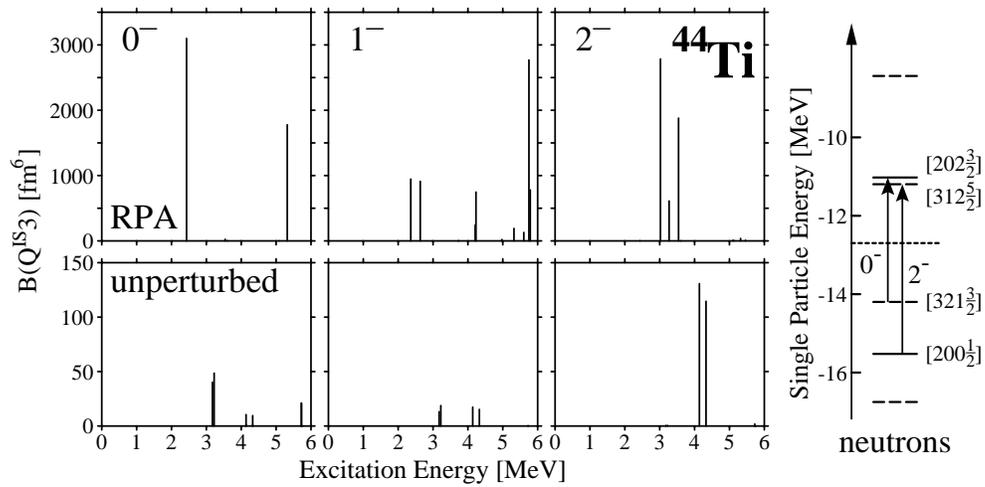}
\caption{{\small 
The same as Fig.~\ref{RPA_unpert.NZ.SIII.SD.prolate},
but for octupole excitations on the triaxial SD state in $^{44}$Ti.
Note that, due to the triaxial deformation, $K$ is not a good 
number for single-particle levels.  
The asymptotic quantum numbers $[N n_z \Lambda \Omega]$
beside the pertinent levels in the rightmost figure 
merely indicate their main components.
}}
\label{unpert.44Ti.SIII}
\end{center}
\end{figure}

Next, let us examine the effects of the triaxial deformations 
of the SD mean field on the RPA eigen-modes discussed above.
Figure \ref{rpa.NZ.triaxial} shows the strongly collective excitations 
in $^{40}$Ca and $^{44}$Ti, obtained by the RPA calculation 
taking into account the triaxiality of the SD mean fields. 
Because  $K$ is not a good quantum number due to the $K$ mixing,
the $K$ values denoted on the right-hand sides of individual levels 
indicate the maximum components,
and the transition strengths for these $K$ values 
are indicated beside the arrows in this figure. 
A clear consequence of the triaxiality is that
the $K \neq$0 modes split into doublets. 
This qualitative feature is common to the RPA calculations 
with the SIII, SkM$^*$ and SLy4 interactions.
Concerning the $^{40}$Ca case, however, the small triaxilality 
($\gamma < 10^{\circ}$) 
in the HF mean field might be easily diminished by, e.g.,
the zero-point fluctuation in the $\gamma$ direction 
and/or pairing correlations. 
Therefore, it is not clear whether this splitting of the $1^-$ mode 
is important or not. 
On the other hand, the triaxial deformations for $^{44}$Ti, 
obtained in the SHF calculations with the SIII and SLy4 interactions, 
are significantly larger than those for $^{40}$Ca
(see the $\gamma$ values indicated at the bottom of 
Fig.~\ref{rpa.NZ.triaxial}).
In such cases, it may be interesting to examine the theoretical
calculation against possible experimental signatures of the triaxility.
For instance, the calculation with the SIII interaction 
yields a ``$K^\pi=0^-$''  collective mode at 2.4 MeV 
and a ``$K^\pi=2^-$'' doublet at 3.0 and 3.5 MeV, 
which possess large isoscalar octupole transition strengths of 
27, 24 and 16 W.u., respectively 
(1 W.u. $\simeq$ 115 fm$^6$ for $^{44}$Ti).
In addition, we also obtain another $0^-$  mode at 5.3 MeV, 
two $1^-$ doublets at 2.5 and 4.2 MeV, 
and another $2^-$ doublet at 3.2 MeV, 
which are moderately collective (not shown in Fig.~\ref{rpa.NZ.triaxial} 
but displayed in Fig.~\ref{unpert.44Ti.SIII}).

The RPA transition strengths are compared 
with the unperturbed particle-hole strengths 
in Fig.~\ref{unpert.44Ti.SIII}.
For the $K=0$ transitions, small unperturbed strengths (about 0.15 W.u.) 
associated with the particle-hole excitations of protons and neutrons 
from the $[321\frac{3}{2}]$ level to the $[202\frac{3}{2}]$ level
are seen at about 3.2 MeV. 
For the $K=2$ transitions,
the peaks in the 4.1-4.3 MeV region in
the unperturbed strengths are associated with 
the particle-hole excitations of protons and neutrons 
from the $[200\frac{1}{2}]$  level to the $[312\frac{5}{2}]$ level.
These ``$K^\pi=0^-$'' and ``$K^\pi=2^-$'' RPA modes are generated 
by collective superpositions of
not only these relatively low-lying configurations but also many
particle-hole configurations lying in the higher energy region.
Thus, the present RPA calculation suggests that 
the appearance of the $2^-$ doublet may serve as a good indicator 
of the triaxial nature of the SD state of $^{44}$Ti.
Experimental search for such a doublet on top of the SD band head 
seems very interesting.

\section{Results of calculation for the neutron-rich Sulfur region}

\begin{figure}[p]
\begin{center}
\includegraphics[width=0.75\textwidth,keepaspectratio]
{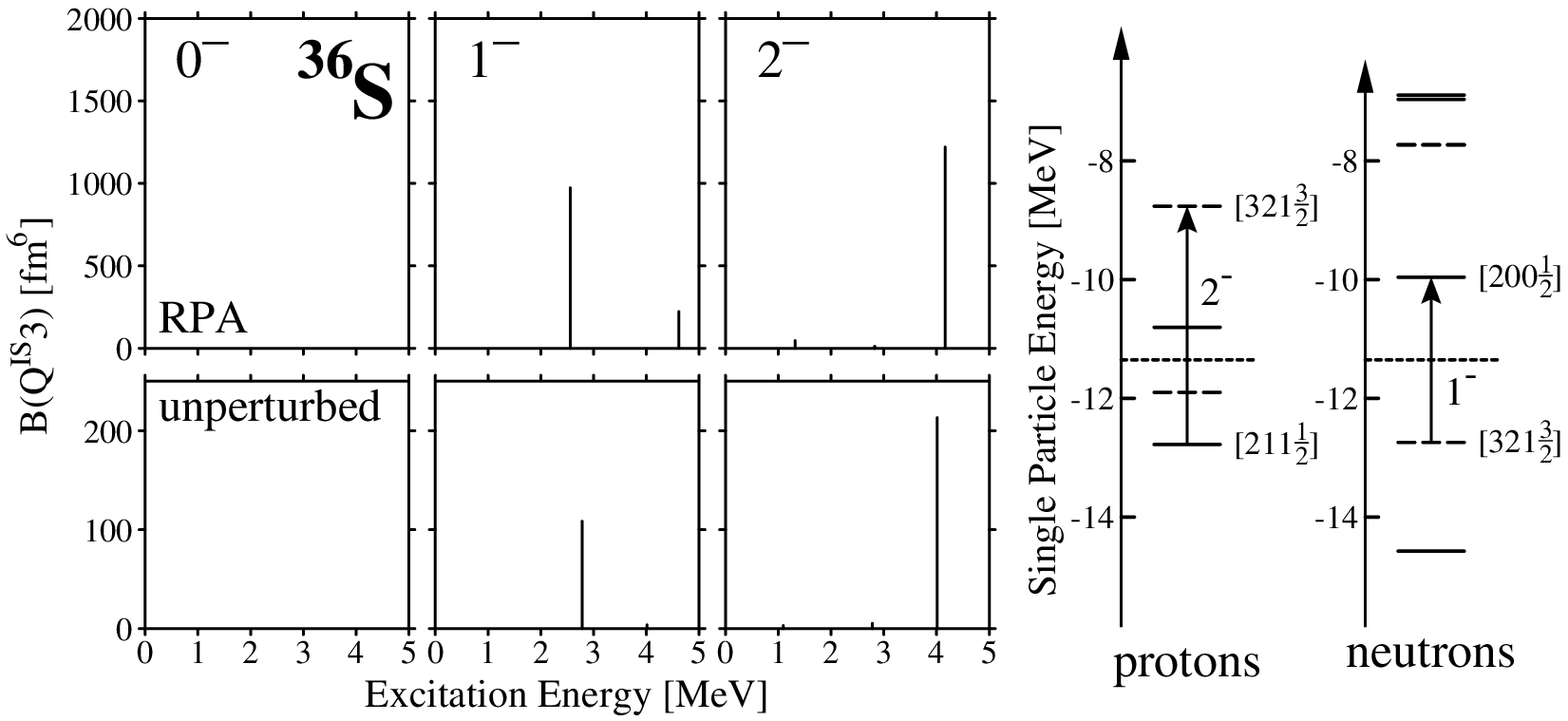}\\
\includegraphics[width=0.75\textwidth,keepaspectratio]
{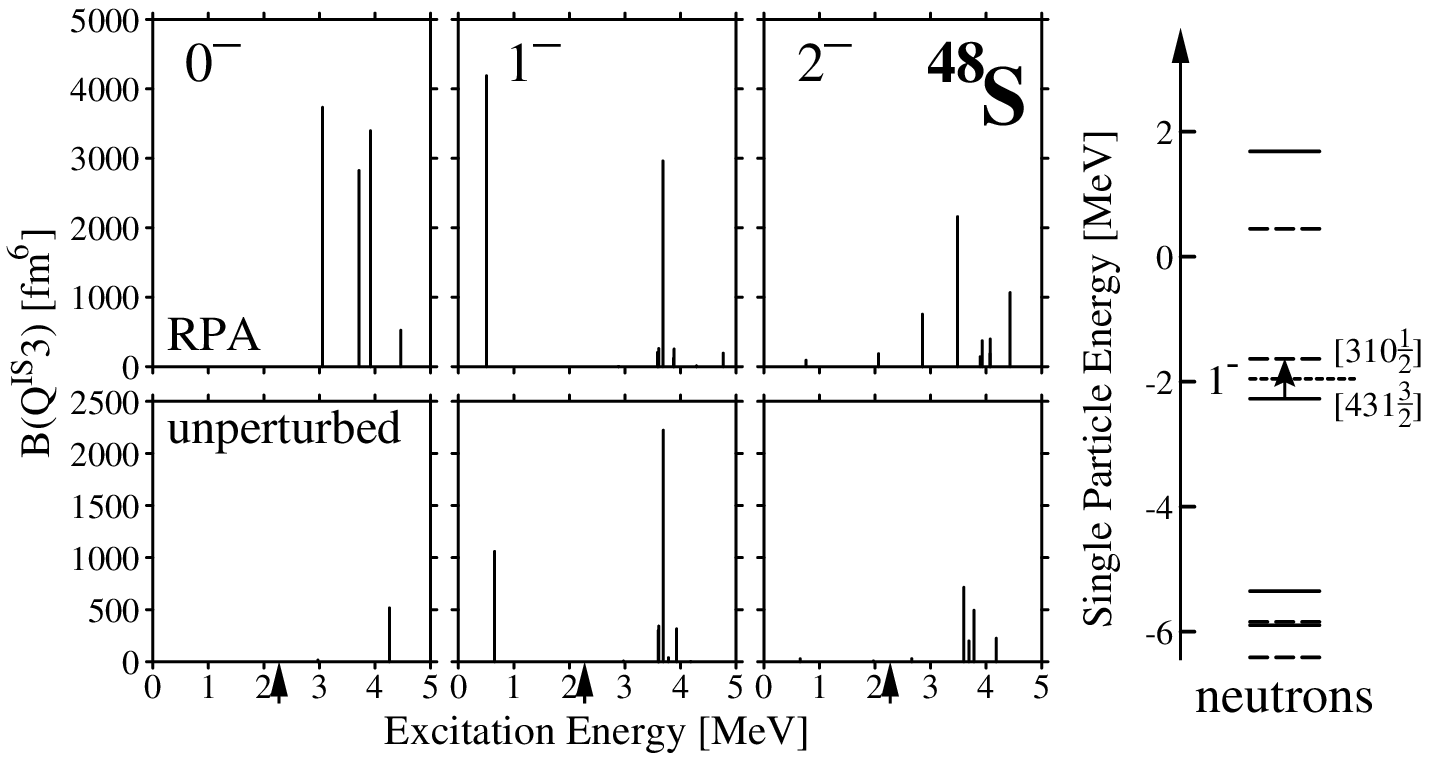}\\
\includegraphics[width=0.75\textwidth,keepaspectratio]
{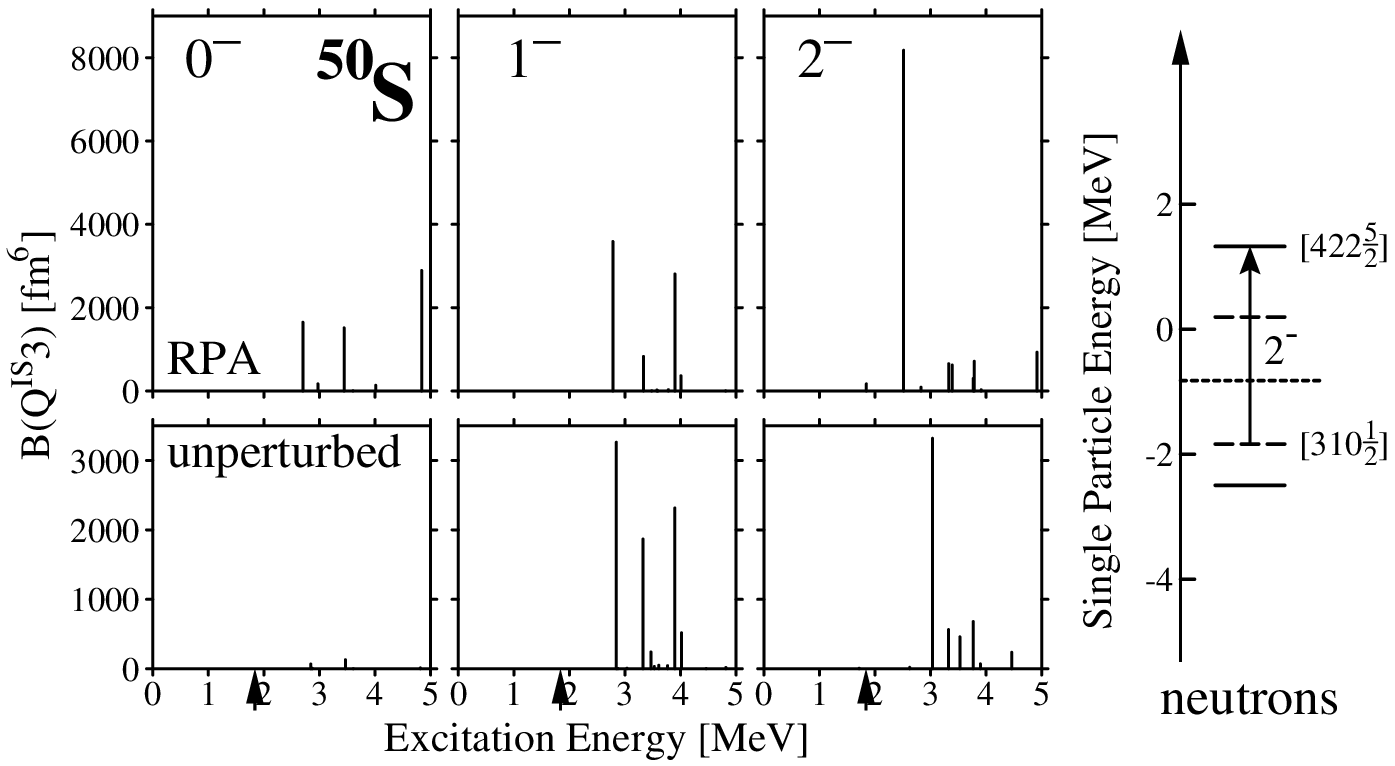}\\
\caption{{\small
Isoscalar octupole transition strengths for $K^{\pi}=0^{-},1^{-}$ and $2^{-}$ 
excitations on the SD states in (a) $^{36}$S, (b) $^{48}$S, and (c) $^{50}$S, 
obtained by the SHF-RPA calculation with the SIII interaction.
The unit is fm$^6$.
For comparison with the RPA strengths, unperturbed particle-hole strengths 
are also shown in the lower panels for each nucleus.
The arrows indicate the threshold energies for neutron emission.
There are no significant strengths for $K^{\pi}=3^{-}$ excitations 
in this low-energy region, so that they are not shown.
Some particle-hole excitations 
near the Fermi surface are drawn by arrows with their $K^{\pi}$ values 
in the rightmost part for each nucleus.
The solid, dashed, and dotted lines indicate the positive-parity levels, 
negative-parity levels, and the Fermi surface, respectively.
The asymptotic quantum numbers 
$[N n_z \Lambda \Omega]$ are indicated for pertinent levels.
For the  positive energy region, only the discretized continuum 
levels possessing resonance character are drawn.
}}
\label{RPA_unpert.S.SIII.SD}
\end{center}
\end{figure}

Figure \ref{RPA_unpert.S.SIII.SD}  
displays the transition strength distributions obtained by 
the SHF-RPA calculation with the SIII interaction 
for low-frequency octupole excitations built on the SD states 
in $^{36,48,50}$S.
Let us first discuss the result of calculation for the neutron-rich $^{36}$S.
For this nucleus, we obtain a $K^\pi=1^-$ mode at 2.6 MeV  
and a $K^\pi=2^-$ mode at 4.2 MeV. 
As they possess the isoscalar octupole transition strengths 
of 13 W.u. and 16 W.u., respectively, 
they are moderately collective 
(1 W.u. $\simeq$ 77 fm$^6$ for $^{36}$S).
It is interesting to compare this result with those for
$^{32}$S and $^{40}$Ca discussed in the previous section.
The $1^-$ mode in this nucleus is similar to 
the $1^-$ mode in $^{40}$Ca,
but its transition strength is less than half of the latter.
This reduction of the collectivity is understood in terms of 
the weakening of the coherence between the proton and neutron excitations 
when one goes from the $N=Z$ nucleus to the $N \ne Z$ nucleus.
Looking into the microscopic details, we find that, for example, 
the particle-hole excitations of protons and neutrons
from the $[321\frac{3}{2}]$ level to the $[200\frac{1}{2}]$ level 
act coherently to produce the $K^\pi=1^-$ collective mode in $^{40}$Ca, 
but this proton excitation is absent in the case of $^{36}$S 
because the $[321\frac{3}{2}]$ level is unoccupied.
Analogous argument also applies to the $K^\pi=2^-$ mode.
This mode is similar to the $2^-$ mode in $^{32}$S,
but, its transition strength is about half of the latter.
This is understood again in terms of the weakening of the coherence 
between the proton and neutron excitations.
For instance, the particle-hole excitations of protons and neutrons
from the $[211\frac{1}{2}]$ level to the $[321\frac{3}{2}]$ level 
act coherently to produce the $K^\pi=2^-$ collective mode in $^{32}$S, 
but this neutron excitation is absent in the case of $^{36}$S 
because the $[321\frac{3}{2}]$ level is already occupied.

Next, let us discuss the result of calculation for $^{48}$S and $^{50}$S
close to the neutron-drip line.
For $^{48}$S, we obtain a $K^\pi=1^-$ mode 
at very low excitation energy, 0.51 MeV.
This mode possesses a large octupole transition strength of 
31 W.u. (1 W.u. $\simeq$ 137 fm$^6$ for $^{48}$S).
The RPA eigen-energy is shifted down 
only slightly (0.14 MeV) from the unperturbed energy (0.65 MeV) of
the neutron excitation from the $[431\frac{1}{2}]$ level 
to the $[310\frac{1}{2}]$ level.  
The unperturbed transition strength for this single-particle 
excitation is extremely large, i.e., $B(Q^\mathrm{IS}3)=7.7$ W.u. 
The major cause of this remarkable strength is understood as follows:  
Both the $[431\frac{1}{2}]$ and $[310\frac{1}{2}]$ levels are 
loosely bound with binding energy about 2 MeV. 
Because, as is well known, wave functions of the loosely bound states 
are significantly extended outside of the half-density radius,
the matrix element of the octupole operator 
between these configurations acquires a large contribution from 
the region outside of the nuclear surface. 
Thus, such a transition strength can become very large  
in nuclei close to the neutron drip.
Although the increase of the RPA transition strength
from the unperturbed strength indicates the presence
of collective effects, it may be appropriate to consider that
the major character of this mode is of single-particle-type. 
In contrast to the low energy peak at 0.51 MeV discussed above, 
the peaks above the neutron-emission threshold 
(indicated by arrows in Fig.~\ref{RPA_unpert.S.SIII.SD}) 
are considered to be associated with the excitations to discretized 
non-resonant continuum states. We examined that these peaks in the 
continuum indeed moves when we increase the box size 
in the numerical calculation. 
Therefore, we cannot give definite physical meanings to these peaks.

For $^{50}$S, we obtain a $K^\pi=2^-$ mode at 2.5 MeV.
Like the $K^\pi=1^-$ mode in $^{48}$S, it is essentially
a single-particle-type excitation, although it has extremely 
large octupole transition strength, 55 W.u. 
(1 W.u. $\simeq$ 149 fm$^6$ for $^{50}$S). 
Namely, it is created mainly by the neutron excitation 
from the loosely bound  $[310\frac{1}{2}]$ level to the 
resonance $[422\frac{5}{2}]$ level. 
The RPA eigen-energy of this mode is shifted down only slightly 
(0.5 MeV) from the unperturbed excitation energy (3.0 MeV) 
of this particle-hole configuration.
Because the resonance $[422\frac{5}{2}]$ level is situated
near the centrifugal barrier top, its wave function
is significantly extended far out side of the half-density radius 
of this nucleus.
Thus, the unperturbed octupole transition strength
for this particle-hole excitation takes an
extremely large value, $B(Q^\mathrm{IS}3)=22$ W.u. 
This resonance interpretation of the $[422\frac{5}{2}]$ level 
is largely based on the analysis made in Ref.~\cite{yos05}
where the deformed Woods-Saxon potential simulating our SHF potential 
was constructed and properties of the discretized continuum states 
for this potential are analysed in detail. It was then found that
the energy and the root-mean-square radius 
of this level were stable against the variation of the box size.
In the same paper, this conclusion was confirmed also 
by carrying out the eigenphase analysis for this level.  
The appearance of this resonance level is  easily understood as 
due to the rather high centrifugal barrier 
for the relatively large value of the symmetry axis component 
of the angular momentum, $\Omega=5/2$. 
We confirmed that this $2^-$ peak at 2.5 MeV does not move 
but the other peaks moves 
when we increase the box size in the numerical calculation. 
Thus, like in $^{48}$S, except for the $2^-$ peak, 
the peaks above the the neutron-emission threshold  
are considered to be associated with the excitations to discretized 
non-resonant continuum states. 

\begin{figure}[p]
\begin{center}
\includegraphics[width=0.80\textwidth,keepaspectratio]
{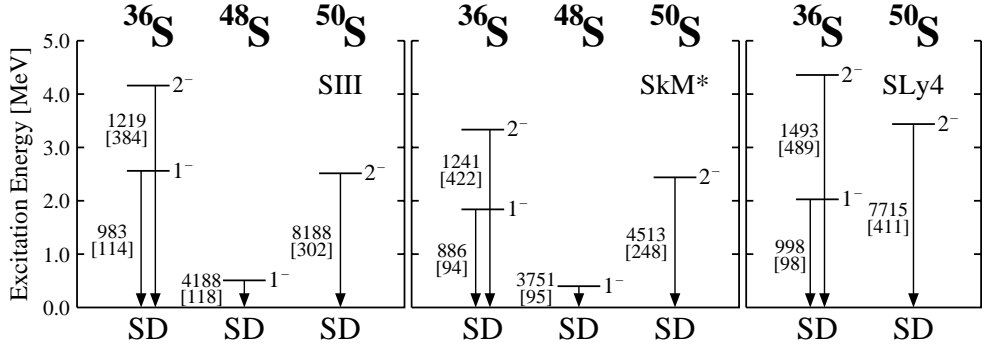}\\
\caption{{\small
Comparison of the SHF-RPA calculations with the SIII, SkM$^*$ and SLy4 
interactions for low-frequency $K^{\pi}=1^{-}$ and $2^{-}$ excitations 
on the SD states in $^{36,48,50}$S.
For SLy4, the SD local minimum does not appear in $^{48}$S.
Only excitations having $B(Q^{\rm IS}3)$ 
greater than 10 W.u. are displayed here 
(see Fig.~\ref{RPA_unpert.S.SIII.SD} 
for other excitations having smaller strengths).
The numbers on the right-hand sides of individual levels 
indicate the $K$ quantum numbers.
The numbers beside the arrows indicate the intrinsic transition matrix 
elements squared of the isoscalar octupole operator, $B(Q^\mathrm{IS}3)$, 
in units of fm$^6$.
Those for the electric octupole operator,
$B(E3)$, are also given in brackets in units of $e^2$fm$^6$.
}}
\label{rpa.S}
\end{center}
\end{figure}

We also carried out the same kind of calculation 
using the SkM$^\ast$\cite{bar82} and the SLy4\cite{cha97} interactions.
The results of calculations with different versions of the
Skyrme interaction are compared in Fig.~\ref{rpa.S}. 
For the SLy4 interaction, the SD local minimum does not appear in $^{48}$S,
so that the RPA calculation was not done for this nucleus. 
Except for this, these Skyrme interactions yield qualitatively similar results 
for the $K^\pi=1^-$ and  $K^\pi=2^-$ modes. 

To accurately describe the spatially extended wave functions,
like the loosely bound levels and resonance levels 
in $^{48}$S and $^{50}$S discussed above,
it is certainly desirable to use a bigger box in the numerical calculation.
Although it is difficult to significantly enlarge the box size 
under the present situation of computing power, 
we plan to manipulate this problem by incorporating 
the adaptive coordinate method into our scheme of numerical calculation. 
We believe, however, that the qualitative features of the present 
RPA calculation will remain valid.
We have not obtained well-developed collective octupole modes 
for $^{48}$S and $^{50}$S. 
In such unstable nuclei close to the neutron-drip line,
there is almost no bound state above the Fermi surface and
neutrons are excited to the continuum states. 
It seems difficult to develop collective correlations among
different particle-hole configurations under such situations.
Quite recently, however, one of the authors (M. Y.) pointed out
that the pairing correlations play an important role for the
emergence of low-frequency collective modes in such drip-line nuclei
\cite{yam05}.
For studying such pairing effects for collective excitations, 
it will be interesting to extend the mixed representation RPA based 
on the SHF mean field to that based on the SHF-Bogoliubov mean field.

\section{Conclusions}

By means of the mixed representation RPA based on the SHF mean field, 
we have investigated low-frequency octupole excitations 
on the SD states in the $N=Z$ nuclei around $^{40}$Ca 
and the neutron-rich Sulfur isotopes. 
The RPA calculation has been carried out on the 3D Cartesian mesh 
in a box, and yielded a number of low-frequency octupole vibrations 
built on the SD states in $^{32}$S, $^{36}$Ar, $^{40}$Ca and $^{44}$Ti. 
In particular, a strongly collective $K^\pi=1^-$ octupole 
vibration has been suggested to appear on top of the SD state in $^{40}$Ca.
For $^{48,50}$S close to the neutron drip line, 
we have found that the low-lying state created by the excitation 
of a single neutron from a loosely bound low $\Omega$ level
to a high $\Omega$ resonance level acquires an extremely strong 
octupole transition strength due to the spatially very extended 
structure of the particle-hole wave functions.

\section*{Acknowledgements}

This work was done as a part of the Japan-US Cooperative Science Program
``Mean-Field Approach to Collective Excitations in Unstable
Medium-Mass and Heavy Nuclei" during the academic years 2003-2004,
and we greatly appreciate useful discussions with the members of this project. 
This work was supported 
by the Grant-in-Aid  for Scientific Research (No. 16540249) 
from the Japan Society for the Promotion of Science.
We also thank the Yukawa Institute for Theoretical Physics at
Kyoto University: Discussions during the YITP workshop YITP-W-05-01 on
"New Developments in Nuclear Self-Consistent Mean-Field Theories" were
useful to complete this work.
The numerical calculations were performed on the NEC SX-5 supercomputers 
at RCNP, Osaka University, 
and at Yukawa Institute for Theoretical Physics, Kyoto University.

\appendix

\section{Explicit expressions of the RPA Hamiltonian}

Here, using the notations of Ref.~\cite{ben02}, 
we recapitulate the quantities appearing in the RPA Hamiltonian 
(\ref{RPA_Hamiltonian}).
More detailed expressions are given in Ref.~\cite{ima03a}. 

The RPA Hamiltonian $h^{(\pm)\lambda}_\RPA$ is written 
\begin{eqnarray}
h^{(\pm)\lambda}_\RPA(\vr\sigma\tau,\vr\sigma'\tau') =
\left[ F^{(\pm)\lambda}_{00}(\vr) \delta_{\tt} 
+ \sum_{\tz} F^{(\pm)\lambda}_{1\tz}(\vr)  \tau^{\tz}_\tt \right]  
\delta_\ss \nonumber \\ 
\quad + \left[ \vect{G}^{(\pm)\lambda}_{00}(\vr) \delta_{\tt} 
+ \sum_{\tz} \vect{G}^{(\pm)\lambda}_{1\tz}(\vr)  
\tau^{\tz}_\tt \right] \cdot \vect{\sigma}_\ss , \\
F^{(\pm)\lambda}_\ttz =
- \nabla \cdot \left[ M^{(\pm)\lambda}_\ttz  \nabla \right] 
+ U^{(\pm)\lambda}_\ttz
- \frac{i}{2} \left[ \nabla \cdot \vect{I}^{(\pm)\lambda}_\ttz 
+ \vect{I}^{(\pm)\lambda}_\ttz \cdot \nabla \right] 
+ U^{(\pm)\lambda}_{\mbox{{\tiny Coul}},\ttz}  \delta_{\pi\tau} ,\\
\vect{G}^{(\pm)\lambda}_\ttz =
- \sum_{\mu}  \nabla_{ \mu} \cdot \left[ \vect{C}^{(\pm)\lambda}_\ttz  
\nabla_{\mu} \right] + \vect{\Sigma}^{(\pm)\lambda}_\ttz
- \frac{i}{2} \sum_{\mu\nu} 
\left[ \nabla_{\mu} B^{(\pm)\lambda}_{\ttz,\mu\nu} 
+  B^{(\pm)\lambda}_{\ttz,\mu\nu}  \nabla_{\mu} \right] \vect{e}_\nu .
\end{eqnarray}
The quantities, $M^{(\pm)\lambda}_\ttz, U^{(\pm)\lambda}_{\ttz}$, {\it etc.} 
appearing in the above expressions are obtained by 
the second derivative of the time-even and time-odd energy functionals, 
$\Ee$ and $\Eo$, with respect to the local densities:
\begin{eqnarray}
M^{(\pm)\lambda}_\ttz &=&
\frac{\partial^2 \Ee_t}{\partial \tau_\ttz \partial \rho_\ttz} 
\rho^{(\pm)\lambda}_\ttz  \ , \\
U^{(\pm)\lambda}_{00} &=& 
\frac{\partial}{\partial \rho_{00}} \left[ \sum_{\ttz} 
\left( \frac{\partial \mathcal{E}_{\mathrm{Sk}}}{\partial \rho_\ttz} 
\rho^{(\pm)\lambda}_\ttz  + \frac{\partial \Eo_t}{\partial \vrho_\ttz} 
\cdot \vrho^{(\pm)\lambda}_\ttz \right)
+\frac{\partial \Ee_0}{\partial \tau_{00}} \tau^{(\pm)\lambda}_{00} 
+ \frac{\partial \Ee_0}{\partial \vect{J}_{ 00}} 
\cdot \vect{J}^{(\pm)\lambda}_{ 00} \right] ,  \\
U^{(\pm)\lambda}_{1\tz} &=& \frac{\partial}{\partial \rho_{1\tz}} 
\left[\frac{\partial \Ee_1}{\partial \rho_{1\tz}} \rho^{(\pm)\lambda}_{1\tz} 
+ \frac{\partial \Ee_1}{\partial \tau_{1\tz}} \tau^{(\pm)\lambda}_{1\tz}
+ \frac{\partial \Ee_1}{\partial \vect{J}_{ 1\tz}} 
\cdot \vect{J}^{(\pm)\lambda}_{ 1\tz} + \frac{\partial \Ee_1}
{\partial \rho_{00}} \rho^{(\pm)\lambda}_{00} \right] \ , \\
B^{(\pm)\lambda}_{\ttz,\mu\nu} &=&
\frac{\partial}{\partial J_{\ttz,\mu\nu}} \left[ \frac{\partial \Ee_t}
{\partial J_{\ttz,\mu\nu}} J^{(\pm)\lambda}_{\ttz,\mu\nu} 
+ \frac{\partial \Ee_t}{\partial \rho_\ttz} 
\rho^{(\pm)\lambda}_\ttz \right] \ , \\
\vect{C}^{(\pm)\lambda}_\ttz &=&
\frac{\partial^2 \Eo_t}{\partial \vect{T}_\ttz \partial 
\vrho_\ttz} \cdot \vrho^{(\pm)\lambda}_\ttz \ , \\
\vect{\Sigma}^{(\pm)\lambda}_\ttz &=&
\frac{\partial}{\partial \vrho_\ttz} \left[
\frac{\partial \Eo_t}{\partial \vrho_\ttz} \cdot \vrho^{(\pm)\lambda}_\ttz 
+ \frac{\partial \Eo_t}{\partial \vect{T}_\ttz} 
\cdot \vect{T}^{(\pm)\lambda}_\ttz
+ \frac{\partial \Eo_t}{\partial \vect{j}_\ttz} 
\cdot \vect{j}^{(\pm)\lambda}_\ttz + \frac{\partial \Eo_t}
{\partial \rho_{00}} \rho^{(\pm)\lambda}_{00} \right] \ , \\
\vect{I}^{(\pm)\lambda}_\ttz &=&
\frac{\partial}{\partial \vect{j}_\ttz} \left[
\frac{\partial \Eo_t}{\partial \vect{j}_\ttz} \cdot 
\vect{j}^{(\pm)\lambda}_\ttz + \frac{\partial \Eo_t}
{\partial \vrho_\ttz} \cdot \vrho^{(\pm)\lambda}_\ttz \right] \ , \\
U^{(\pm)\lambda}_{\mbox{{\tiny Coul}},\ttz} &=&
\frac{\partial}{\partial \rho_\ttz} \left[ \frac{\partial 
\mathcal{E}_{\mbox{{\tiny Coul}}}}{\partial \rho_{00}} 
\rho^{(\pm)\lambda}_{00}
+ \frac{\partial \mathcal{E}_{\mbox{{\tiny Coul}}}}{\partial 
\rho_{10}} \rho^{(\pm)\lambda}_{10} \right] . 
\end{eqnarray}
The local transition densities $\rho^{(\pm)\lambda}_\ttz$, 
the local transition spin densities $\vrho^{(\pm)\lambda}_\ttz$, 
the local transition kinetic energy densities $\tau^{(\pm)\lambda}_\ttz$, 
the local transition kinetic spin densities 
$\vect{T}^{(\pm)\lambda}_\ttz$, 
the local transition current densities $\vect{j}^{(\pm)\lambda}_\ttz$, 
the local transition spin-orbit current tensors 
$J^{(\pm)\lambda}_{\ttz,\mu\nu}$ 
are given by 
\begin{eqnarray}
\rho^{(\pm)\lambda}_\ttz(\vr) &=& \rho^{(\pm)\lambda}_\ttz(\vr,\vr), \\
\vrho^{(\pm)\lambda}_\ttz(\vr) &=& \vrho^{(\pm)\lambda}_\ttz(\vr,\vr), \\
\tau^{(\pm)\lambda}_\ttz(\vr) &=& \left. \nabla \cdot \nabla' 
\rho^{(\pm)\lambda}_\ttz(\vr,\vr') \right|_{\vr=\vr'}, \\
\vect{T}^{(\pm)\lambda}_\ttz(\vr) &=& \left. \nabla \cdot \nabla' 
\vrho^{(\pm)\lambda}_\ttz(\vr,\vr') \right|_{\vr=\vr'}, \\
\vect{j}^{(\pm)\lambda}_\ttz(\vr) &=& - \frac{i}{2}\left. 
\left( \nabla - \nabla' \right) \rho^{(\pm)\lambda}_\ttz(\vr,\vr') 
\right|_{\vr=\vr'}, \\
J^{(\pm)\lambda}_{\ttz,\mu\nu}(\vr) &=& - \frac{i}{2} \left. 
\left( \nabla - \nabla' \right)_\mu \vrho^{(\pm)
\lambda}_{\ttz,\nu}(\vr,\vr') \right|_{\vr=\vr'}.
\end{eqnarray}
The local transition spin-orbit currents 
$\vect{J}^{(\pm)\lambda}_\ttz$ are defined by
\begin{eqnarray}
\vect{J}^{(\pm)\lambda}_\ttz &=& \sum_{\mu\nu\omega} 
\epsilon_{\mu\nu\omega} J^{(\pm)\lambda}_{\ttz,\nu\nu}\vect{e}_\omega,
\end{eqnarray}
where $\epsilon_{\mu\nu\omega}$ is the Levi-Civita symbol and 
$\vect{e}_\omega$ is the unit vector in the $\omega$-direction.
The scalar transition densities $\rho^{(\pm)\lambda}_\ttz (\vr,\vr')$ 
and the vector transition densities $\vrho^{(\pm)\lambda}_\ttz (\vr,\vr')$
appearing in the above equations are defined by
decomposing the transition density matrix 
$\rho^{(\pm)\lambda}(x,x') = 
\rho^{(\pm)\lambda}(\vr\sigma\tau, \vr'\sigma'\tau')$
into the spin-isospin channels:
\begin{eqnarray}
\rho^{(\pm)\lambda}(\vr\sigma\tau, \vr'\sigma'\tau') 
=&& \frac{1}{4}\bigg[ \rho^{(\pm)\lambda}_{00}(\vr,\vr') \delta_\ss 
\delta_\tt + \vrho^{(\pm)\lambda}_{00}(\vr,\vr') \cdot \vect{\sigma}_\ss 
\delta_\tt \nonumber \\
&& + \delta_\ss  \sum^1_{\tz=-1} \rho^{(\pm)\lambda}_{1\tz}(\vr,\vr') 
\tau^\tz_{\tau\tau'}
+ \sum^1_{\tz=-1} \vrho^{(\pm)\lambda}_{1\tz}(\vr,\vr') 
\cdot \vect{\sigma}_\ss \tau^\tz_{\tau\tau'}\bigg].
\end{eqnarray}
In the above expressions, the charge-exchange $\tz=\pm 1$ components 
are included for completeness, although they do not contribute
to the excitation modes considered in this paper.


\newpage
\begin{table}
\caption{Classification of octupole and dipole operators
according to the parity quantum numbers with respect to 
reflections about the $(y,z)$-, $(z,x)$- and $(x,y)$-planes.
\label{table:symmetry}
}
\begin{center}
\begin{tabular}{|r|r|r|} \hline
~symmetry~ & ~octupole~ & ~dipole~ \\ \hline
$(-,+,+)$~ & ~$Q_{31}^{(+)}$, ~$Q_{33}^{(+)}$~& ~$x$~~~~ \\ \hline
$(+,-,+)$~ & ~$Q_{31}^{(+)}$, ~$Q_{33}^{(+)}$~& ~$y$~~~~ \\ \hline
$(+,+,-)$~ & ~$Q_{30}$, ~$Q_{32}^{(+)}$~ & ~$z$~~~~ \\ \hline
$(-,-,-)$~ & ~$Q_{32}^{(-)}$~~~~ & none~~~\\ \hline
\end{tabular}
\end{center}
\end{table}
\newpage


\newpage

\newpage

\newpage

\newpage

\end{document}